\documentclass[5p, sort&compress]{elsarticle}

\usepackage{lineno,hyperref}
\modulolinenumbers[5]
\journal{Acta Materialia}

	\usepackage{subfigure}
	\usepackage{booktabs}
	\usepackage{multirow}
	\usepackage{rotating}
	\usepackage{longtable}
	\usepackage{setspace}
	\usepackage{esvect} 
	\usepackage{morefloats}
	\usepackage{arydshln}
	\usepackage{setspace}
	\usepackage{tabularx}
	\usepackage{lmodern}
	\usepackage{lscape}
	\DeclareUnicodeCharacter{2212}{-}

\usepackage{color}
\usepackage{amsmath}
\usepackage{amssymb}
\usepackage{graphicx}
\usepackage{siunitx}
	\DeclareSIUnit{\atom}{at.}
	\DeclareSIUnit{\weight}{wt.}
	\DeclareSIUnit{\oersted}{Oe}
	\DeclareSIUnit{\emu}{emu}
	\DeclareSIUnit{\fe}{Fe}
\usepackage{subfigure}
\begin{document}
\begin{frontmatter}
\author[label1]{Andreas Taubel\corref{cor1}}
\ead{andreas.taubel@tu-darmstadt.de}
\author[label1]{Benedikt Beckmann\corref{cor1}}
\ead{benedikt.beckmann@tu-darmstadt.de}
\author[label1]{Lukas Pfeuffer}
\author[label1]{Nuno Fortunato}
\author[label1]{Franziska Scheibel}
\author[label1]{Semih Ener}
\author[label2]{Tino~Gottschall}
\author[label1]{Konstantin P. Skokov}
\author[label1]{Hongbin Zhang}
\author[label1]{Oliver Gutfleisch}
\cortext[cor1]{Corresponding authors: A. Taubel and B. Beckmann contributed equally to the manuscript.}
\address[label1]{Institute for Materials Science, Technical University of Darmstadt, 64287 Darmstadt,
Germany}
\address[label2]{Dresden High Magnetic Field Laboratory (HLD-EMFL), Helmholtz-Zentrum Dresden-Rossendorf, 01328 Dresden, Germany}
\title{Tailoring magnetocaloric effect in all-$d$-metal \mbox{Ni-Co-Mn-Ti} Heusler alloys: a combined experimental and theoretical study}
\begin{abstract}

Novel Ni-Co-Mn-Ti all-$d$-metal Heusler alloys are exciting due to large multicaloric effects combined with enhanced mechanical properties. An optimized heat treatment for a series of these compounds leads to very sharp phase transitions in bulk alloys with isothermal entropy changes of up to \SI{38}{\joule\per\kilo\gram\per\kelvin} for a magnetic field change of \SI{2}{\tesla}. The differences of as-cast and annealed samples are analyzed by investigating microstructure and phase transitions in detail by optical microscopy. We identify different grain structures as well as stoichiometric (in)homogenieties as reasons for differently sharp martensitic transitions after ideal and non-ideal annealing. We develop alloy design rules for tuning the magnetostructural phase transition and evaluate specifically the sensitivity of the transition temperature towards the externally applied magnetic fields ($\frac{dT_t}{\mu_0dH}$) by analyzing the different stoichiometries. We then set up a phase diagram including martensitic transition temperatures and austenite Curie temperatures depending on the $e/a$ ratio for varying Co and Ti content. The evolution of the Curie temperature with changing stoichiometry is compared to other Heusler systems. Density Functional Theory calculations reveal a correlation of $T_C$ with the stoichiometry as well as with the order state of the austenite. This combined approach of experiment and theory allows for an efficient development of new systems towards promising magnetocaloric properties. Direct adiabatic temperature change measurements show here the largest change of \SI{-4}{\kelvin} in a magnetic field change of \SI{1.93}{\tesla} for Ni$_{35}$Co$_{15}$Mn$_{37}$Ti$_{13}$. 

\end{abstract}


\end{frontmatter}

\section{Introduction}
\label{Intro}

Energy-efficient technologies are necessary to slow down climate change and the depletion of natural energy resources. Since the global power demand for cooling will likely exceed the energy consumption for heating in the second half of this century~\cite{Isaac2009}, it is pivotal to develop new energy-efficient, inexpensive and environmentally friendly cooling technologies, which will result in reduced CO$_2$ emissions~\cite{Gutfleisch2011}.
The magnetocaloric effect can provide a solution with an increased energy-efficiency compared to conventional vapor-compression technology~\cite{Zimm1998, Gutfleisch2011, Franco2012}. 

Various material systems such as La-Fe-Si, Fe-Rh, and Fe$_2$P-type alloys show attractive magnetocaloric properties~\cite{Fujita2003, Gutfleisch2005, Nikitin1990, Tegus2002, Fries2017, Gottschall2019}.  A versatile class is represented by the huge family of \mbox{Ni(-Co)-Mn-X} Heusler alloys, with X being a main group element, e.g. Ga, In, Sn, Sb or Al~\cite{Sutou2004, Krenke2005a, Dubenko2009, LiuNature, Gottschall2015}. They crystallize in the Heusler structure (L2$_1$) with four interpenetrating face-centered cubic sublattices~\cite{Heusler1934}. The sublattices are occupied according to the valency with Ni on the A- and C-sites, Mn on the B-site and the X element on the D-site sublattice. A disorder between B- and D-sites is very common and leads to the B2 structure. Typical for these alloys is the occurrence of a martensitic transition with a tunable transition temperature $T_t$. This transition can be coupled to a magnetic phase transition leading to a first-order phase transition with a conventional or inverse magnetocaloric effect. Since the (local) transition temperature is sensitively depending on the chemical composition, sharpest transitions are obtained for chemically homogeneous compounds. Thus, an optimized sample preparation process is of utmost importance. So far, the most promising Heusler alloy for magnetocaloric purposes is \mbox{Ni-Co-Mn-In} with an adiabatic temperature change $\Delta T_{ad}$ of \SI{-8}{\kelvin} in a magnetic field of \SI{2}{\tesla}~\cite{Gottschall2016}. Since In is an expensive and critical element, it is worth searching for other Heusler alloys with a comparable or even superior performance. 

Recent studies on all-$d$-metal Heusler alloys with X being a transition metal occupying the D-sites of the Heusler lattice showed a magnetostructural phase transition of first-order type for Ni-rich and Mn-rich \mbox{Ni-Co-Mn-Ti} systems~\cite{Wei2015, Wei2016}, where highest isothermal entropy changes of \SI{10}{\joule\per\kilo\gram\per\kelvin} in magnetic fields of \SI{2}{\tesla} are reported for bulk Ni$_{35}$Co$_{15}$Mn$_{35}$Ti$_{15}$~\cite{Wei2015}. An increasing amount of Ti stabilizes the austenitic phase, whereas Co substitution additionally increases the austenitic Curie temperature $T_{C}^{A}$. For a certain amount of Co atoms on the Ni sites, the desired martensitic transformation from ferromagnetic austenite to weak-magnetic martensite is enabled. One drawback of many magnetocaloric compounds is the brittleness leading to mechanical degradation even after a small number of magnetocaloric cycles~\cite{Guillou2014, Bartok2016, Gutfleisch2016, Taubel2017}. The all-$d$-metal Heusler alloys are suggested to solve this problem since the strength of hybridized $d$-$d$ bonding leads to a higher mechanical stability~\cite{Wei2015, Liu2019a} making them very interesting for magnetocaloric~\cite{Wei2015, Wei2016, Liu2019}, barocaloric~\cite{Aznar2019} and elastocaloric~\cite{Wei2019, Yan2019} purposes. Enlarged magnetocaloric effects have been shown recently by isothermal entropy changes of up to \SI{27}{\joule\per\kilo\gram\per\kelvin} in magnetic field changes of \SI{2}{\tesla} for mechanically stable melt-spun ribbons~\cite{Neves-Bez2019}. However, the performance of the all-$d$-metal Heusler alloys has not been reported in terms of direct measurements of the adiabatic temperature change for the magnetocaloric effect. 

We analyze the magnetocaloric effect of the Ni$_\text{50-x}$Co$_\text{x}$Mn$_\text{50-y}$Ti$_\text{y}$ sample series by systematically varying the Co content as well as the Ti content. We use this stoichiometric series to set up a phase diagram. The results are then compared to other Heusler systems and correlated with Density Functional Theory (DFT) calculations, which describe the experimental findings accurately. As already shown in~\cite{Taubel2018}, the annealing conditions need to be chosen carefully for the production of magnetocaloric Heusler alloys. In a systematic study, we optimize the heat treatment and improve the magnetocaloric properties of bulk samples significantly. This underlines that the key to sharp phase transitions is a good stoichiometric homogeneity, which we study by correlating the temperature-dependent magnetization with the respective microstructure for differently annealed samples. Finally, we investigate the magnetocaloric effect for this system by direct measurements of the adiabatic temperature change in order to assess the cyclic performance of all-$d$-metal Heusler alloys for magnetic-field induced cooling applications. 

\begin{figure*}[htb]
    \centering
    \begin{minipage}{0.32\linewidth}
        \centering
        \includegraphics[width=\linewidth]{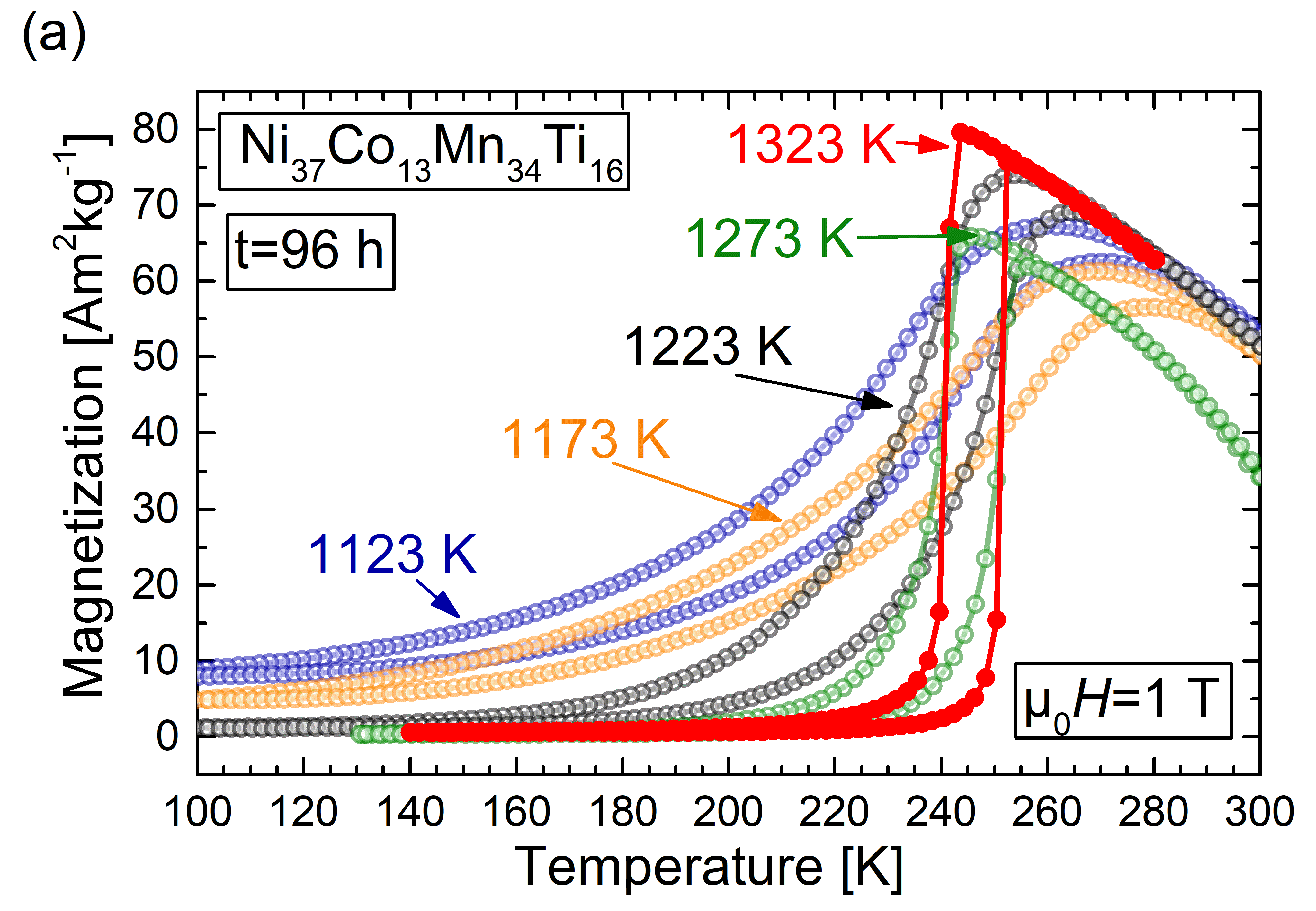}
          \end{minipage}
    \begin{minipage}{0.32\linewidth}
        \centering
        \includegraphics[width=\linewidth]{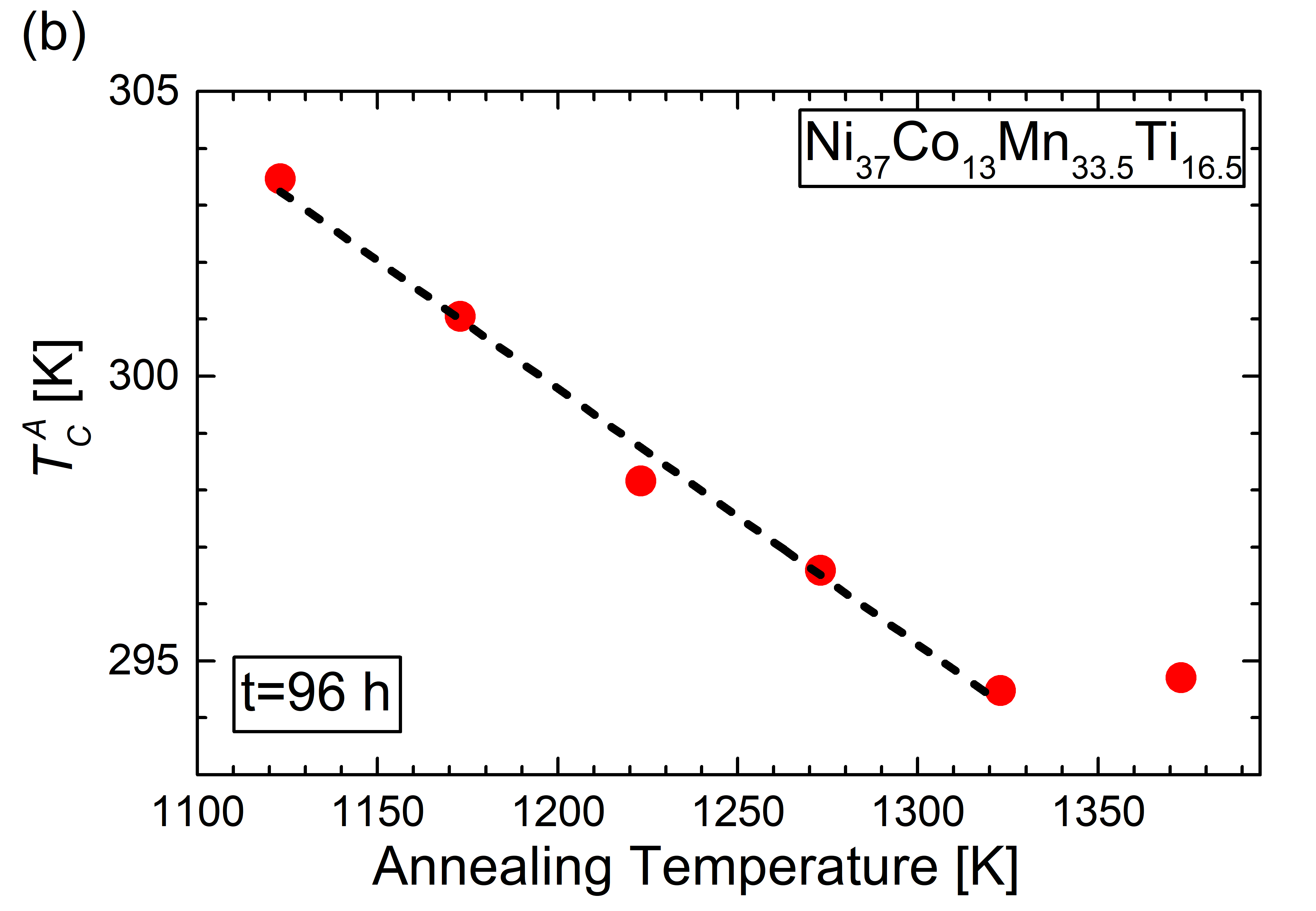}
	\end{minipage}
	\begin{minipage}{0.32\linewidth}
        \centering
        \includegraphics[width=\linewidth]{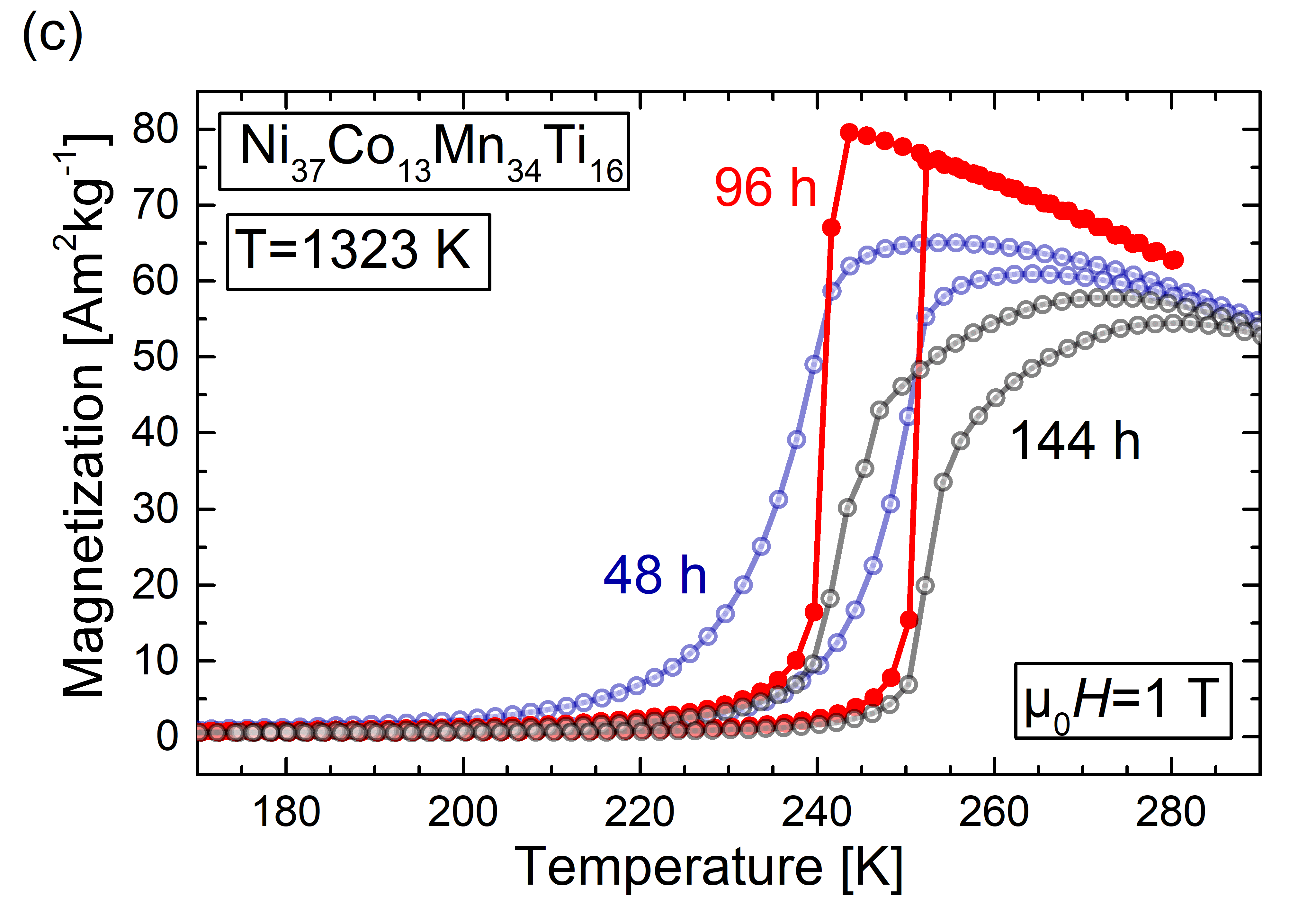}
	\end{minipage}
	\caption{Magnetization as a function of temperature for samples produced under different annealing conditions by varying temperature (a) and duration (c) shown for Ni$_{37}$Co$_{13}$Mn$_{34}$Ti$_{16}$. The evolution of the austenitic Curie temperature for different annealing temperatures for Ni$_{37}$Co$_{13}$Mn$_{33.5}$Ti$_{16.5}$ is shown in (b). The magnetization curves have been measured in a magnetic field of \SI{1}{\tesla}, whereas $T_C^A$ has been determined in a magnetic field of \SI{0.01}{\tesla}. Consequently, the most suitable properties can be obtained by a heat treatment at \SI{1323}{\kelvin} for \SI{96}{\hour} followed by water quenching.}
	\label{annealing}
\end{figure*}
 
\section{Material and Methods}
\label{Experimental}

Nominal Ni$_\text{50-x}$Co$_\text{x}$Mn$_\text{50-y}$Ti$_\text{y}$ compositions were prepared by arc melting in Ar atmosphere for three series, each with fixed Co content ($x=13, 15, 17$) and varying Ti concentration. Due to evaporation losses of Mn during melting, an excess of \SI{3}{\percent} Mn was added. In the following, samples will be denoted by their nominal composition. An overview of the samples with actual stoichiometries determined by energy-dispersive X-ray spectroscopy (EDS) and selected properties from this work are shown in table~\ref{tab:Overview}. The samples were melted six times and turned around after each melting step to ensure a homogeneous melting. For an optimization of the heat treatment conditions, a homogenization process was carried out for the series with a Co content of \SI{13}{\atom\percent} at different temperatures between \SI{1123}{\kelvin} and \SI{1373}{\kelvin} for \SI{96}{\hour} in sealed quartz ampoules in Ar atmosphere. Afterwards, the resulting optimal annealing temperature was kept constant for a further study by varying the duration of the heat treatment from \SI{48}{\hour} to \SI{144}{\hour}. The samples were always quenched subsequently by breaking the tube in water at room temperature. 

Magnetization measurements were performed on a \textit{LakeShore} vibrating sample magnetometer (VSM) 7410 and a Physical Properties Measurement System (PPMS) from \textit{Quantum Design}. 

A \textit{Zeiss Axio Imager.D2m} microscope equipped with polarized light function and a temperature stage has been used for optical microscopy. Secondary electron (SE) and backscatter electron (BSE) imaging was carried out on a \textit{Tescan Vega3} electron microscope equipped with an \textit{EDAX Octane Plus} system for EDS analysis.

The determination of the isothermal entropy change $\Delta s_{T}$ was carried out by temperature-dependent magnetization measurements (isofield protocol). First derivatives of the resulting $M(T)$-curves were determined for numerical integration according to equation~\ref{EquationDeltaSm} resulting from Maxwell's relation~\cite{Oliveira2010}.

\begin{equation}
\Delta{s_{T}}(T,\Delta H)=\int_{H_1}^{H_2} \left(\frac{\partial M(T,H)}{\partial T}\right)_H dH .
\label{EquationDeltaSm}
\end{equation}

The isothermal entropy change was calculated from $M(T)$ curves of bulk samples measured at varying fields from \SI{0.1}{} to \SI{2}{\tesla} with a step size of \SI{0.2}{\tesla}. 

Adiabatic temperature changes were measured directly in a purpose-built device described in detail in~\cite{LiuNature}. The temperature change was measured by an attached thermocouple under sinusoidal field application and removal with a maximum field strength of \SI{1.93}{\tesla}. For the continuous protocol, the sample was heated/cooled with a constant rate and the magnetic field was applied on the run at the measurement temperatures. 

Exchange interaction ($J_{ij}$) calculations and lattice constant optimization were performed in the KKR-Green's function formalism as implemented in the SPR-KKR code version 7.7.3~\cite{Ebert, Ebert2011}. All calculations were carried out using the atomic sphere approximation mode and the exchange correlation functional was approximated based on the PBE parametrization~\cite{Perdew1997}. To guarantee good convergence for $T_C$, the k-point mesh density was set to be 4000 and an angular moment cut-off of $l=$3 was implemented, with a $J_{ij}$ being calculated up to a radius of 4 times the lattice constant. The coherent potential approximation was used to treat the L2$_1$/B2 chemical disorder~\cite{Butler1985}. An in-house Monte Carlo code was used to obtain Curie temperatures based on the Heisenberg Model and the parameterized $J_{ij}$ from SPR-KKR, solved using the Metropolis method~\cite{Metropolis1953}. We adopted a minimum of 40 independently initialized ensembles running in parallel for 15000 Monte Carlo steps (MCS), with thermodynamic amounts being measured every 1000 MCS for decorrelation and 2500 MCS for the initial thermalization at each temperature value. The Curie temperature  was determined by the intersection of the Binder cumulant between systems composed of 10x10x10 and 20x20x20 lattice cells.

\section{Results and Discussion}
\label{Results}

\subsection{Optimization of the heat treatment}
\label{chap:heat treatment}

In order to produce Heusler samples with sharp phase transitions leading to large magnetocaloric effects, the homogenizing heat treatment has been optimized. Figure~\ref{annealing}~(a) shows that the phase transition for a heat treatment for \SI{96}{\hour} at \SI{1173}{\kelvin} and below is very broad for the exemplarily shown sample. This broad transition behavior leads to reduced magnetocaloric effects due to the diminished $\frac{dM}{dT}$. For higher annealing temperatures, the phase transitions of the samples become sharper indicating a better homogenization. This effect saturates at a temperature of \SI{1323}{\kelvin} and the transitions overall get worse for an annealing at \SI{1373}{\kelvin}. The factors considered for the quality of the phase transition are the transition width, the magnetization change of the transition, and the thermal hysteresis. The trend of the austenitic Curie temperature $T_C^A$ plotted for different annealing temperatures in Fig.~\ref{annealing}~(b) points out that a better homogenization with higher annealing temperature is accompanied by a reduction of $T_C^A$. The annealing temperature of \SI{1373}{\kelvin}, which is identified from magnetization data as too high, results in a slight increase of $T_C^A$. Besides atomic ordering effects, the homogenization of the stoichiometric variations present in as-cast state are responsible for this effect, which will be underlined by the microscopical investigations in the subsequent section. This analysis allows us to select an optimal temperature for the homogenization step. Like for other magnetocaloric Heusler alloys, this temperature must be high enough with respect to the melting point in order to enable optimized diffusion processes that result in best stoichiometric homogeneity~\cite{Taubel2018}. The melting point of \mbox{Ni-Co-Mn-Ti} has been determined by a differential scanning calorimetry (DSC) measurement to be at \SI{1380}{\kelvin}. This is around \SIrange{50}{100}{\kelvin} larger than for the \mbox{Ni(-Co)-Mn-Sn} system~\cite{Schlagel2008, Wachtel1983, Taubel2018} being conform with the optimal annealing temperature, which is \SI{100}{\kelvin} larger for \mbox{Ni-Co-Mn-Ti} compared to \mbox{Ni(-Co)-Mn-Sn}. An ideal annealing is thus obtained at temperatures that are approximately \SI{50}{\kelvin} below the melting point. The heat treatment temperature must not be closer than \SI{50}{\kelvin} to the melting point to prevent local melting of the sample and immoderate evaporation of manganese.

In a second step, the optimal annealing temperature is kept constant and the annealing time is now varied. The $M(T)$-curves in Fig.~\ref{annealing}~(c) show that the initial time of \SI{96}{\hour} leads to the best result. A shorter time results in a broader transition due to insufficient homogenization, whereas longer annealing also increases the width of the phase transition. This can be due to an enhanced evaporation of Mn (in analogy to high annealing temperatures) leading to stoichiometric variations in the compound and thus to a broader distribution of local transition temperatures.

Optimal annealing conditions are always an interplay between temperature and duration and defining an ideal set of annealing parameters is a complex issue for such delicate alloys. The optimal annealing for the \mbox{Ni-Co-Mn-Ti} sample series is here found to be at \SI{1323}{\kelvin} for \SI{96}{\hour}. 

\begin{table*}[htb]
\begin{center}
\caption{Nominal compositions of the produced Ni$_\text{50-x}$Co$_\text{x}$Mn$_\text{50-y}$Ti$_\text{y}$ samples with actual compositions (next four columns) determined by EDS together with the corresponding $e/a$ ratio, austenite start temperature  ($A_s$, determined in \SI{1}{\tesla}), width of the transition ($\Delta T_\text{width}$), thermal hysteresis ($\Delta T_\text{hyst}$), shift of the transition temperature with applied magnetic fields ($dT_t/\mu_0dH$, average in fields of \SI{0.1}{\tesla}, \SI{1}{\tesla} and \SI{2}{\tesla}) and austenitic Curie temperature ($T_C^A$, determined in \SI{0.01}{\tesla}).}
\begin{tabular}{ccccccccccc}
\toprule
nominal & Ni & Co & Mn & Ti & $e/a$ & $A_s$ & $\Delta T_\text{width}$ & $\Delta T_\text{hyst}$ & $dT_t/\mu_0dH$ & $T_C^A$ \\
 & [\SI{}{\atom\percent}] & [\SI{}{\atom\percent}] & [\SI{}{\atom\percent}] & [\SI{}{\atom\percent}]  &  & [\SI{}{\kelvin}] & [\SI{}{\kelvin}] & [\SI{}{\kelvin}] & [\SI{}{\kelvin\per\tesla}] & [\SI{}{\kelvin}] \\
\midrule
Ni$_{37}$Co$_{13}$Mn$_{35}$Ti$_{15}$ & 36.6 & 12.0 & 36.3 & 15.1 & 7.885 & 310.1 & 16.0 & 8.4 & -1.2 & 340.2 \\
Ni$_{37}$Co$_{13}$Mn$_{34.5}$Ti$_{15.5}$ & 35.4 & 12.8 & 36.8 & 15.0 & 7.866 & 293.3 & 11.7 & 5.5 & -1.6 & 333.8 \\
Ni$_{37}$Co$_{13}$Mn$_{34}$Ti$_{16}$ & 36.0 & 12.7 & 35.2 & 16.1 & 7.848 & 249.7 & 7.2 & 9.9 & -1.7 & 311.6 \\
Ni$_{37}$Co$_{13}$Mn$_{33.5}$Ti$_{16.5}$ & 35.9 & 12.7 & 34.7 & 16.7 & 7.831 & 225.5 & 14.4 & 11.4 & -2.1 & 294.5 \\
Ni$_{37}$Co$_{13}$Mn$_{33}$Ti$_{17}$ & 34.1 & 13.1 & 36.2 & 16.6 & 7.785 & 121.5 & 33.3 & 20.0 & -7.4 & 303.5 \\
Ni$_{35}$Co$_{15}$Mn$_{39}$Ti$_{11}$ & 33.6 & 14.6 & 40.8 & 11.0 & 7.967 & 485.1 & 23.0 & 19.1 & -0.1 & 480.5 \\
Ni$_{35}$Co$_{15}$Mn$_{38}$Ti$_{12}$ & 33.6 & 14.7 & 39.6 & 12.1 & 7.939 & 432.7 & 28.5 & 28.4 & -1.7 & 462.1 \\
Ni$_{35}$Co$_{15}$Mn$_{37}$Ti$_{13}$ & 33.6 & 14.6 & 38.8 & 13.0 & 7.910 & 346.4 & 42.3 & 17.6 & -2.8 & 442.1 \\
Ni$_{35}$Co$_{15}$Mn$_{36}$Ti$_{14}$ & 33.3 & 14.8 & 37.7 & 14.2 & 7.871 & 278.2 & 22.1 & 14.2 & -3.4 & 414.7 \\
Ni$_{35}$Co$_{15}$Mn$_{35}$Ti$_{15}$ & 33.8 & 14.8 & 36.4 & 15.0 & 7.859 & 211.0 & 32.8 & 17.0 & -5.0 & 387.0 \\
Ni$_{35}$Co$_{15}$Mn$_{34}$Ti$_{16}$ & 33.7 & 14.8 & 35.4 & 16.1 & 7.823 & 136.3 & 31.7 & 20.6 & -9.3 & 358.7 \\
Ni$_{33}$Co$_{17}$Mn$_{40}$Ti$_{10}$ & 31.6 & 16.5 & 41.8 & 10.1 & 7.975 & 509.8 & 22.0 & 30.3 & -1.0 & 535.3 \\
Ni$_{33}$Co$_{17}$Mn$_{39}$Ti$_{11}$ & 31.6 & 16.6 & 40.8 & 11.0 & 7.949 & 454.5 & 29.8 & 35.0 & -3.1 & 517.7 \\
Ni$_{33}$Co$_{17}$Mn$_{38}$Ti$_{12}$ & 31.6 & 16.6 & 39.7 & 12.1 & 7.916 & 378.0 & 43.1 & 41.3 & -5.1 & 498.5 \\
Ni$_{33}$Co$_{17}$Mn$_{37}$Ti$_{13}$ & 31.8 & 16.7 & 38.4 & 13.1 & 7.894 & - & - & - & - & 478.9 \\
Ni$_{33}$Co$_{17}$Mn$_{36}$Ti$_{14}$ & 31.9 & 16.6 & 37.4 & 14.1 & 7.866 & - & - & - & - & 458.2 \\
Ni$_{33}$Co$_{17}$Mn$_{35}$Ti$_{15}$ & 31.8 & 16.6 & 36.4 & 15.2 & 7.832 & - & - & - & - & 435.0 \\

\bottomrule
\end{tabular}
\label{tab:Overview}
\end{center}
\end{table*}

\subsection{Influence of stoichiometry and magnetic field on the phase transitions}
\label{chap:MT}

\begin{figure}[htb]
    \centering
    \begin{minipage}{0.95\linewidth}
        \centering
        \includegraphics[width=\linewidth]{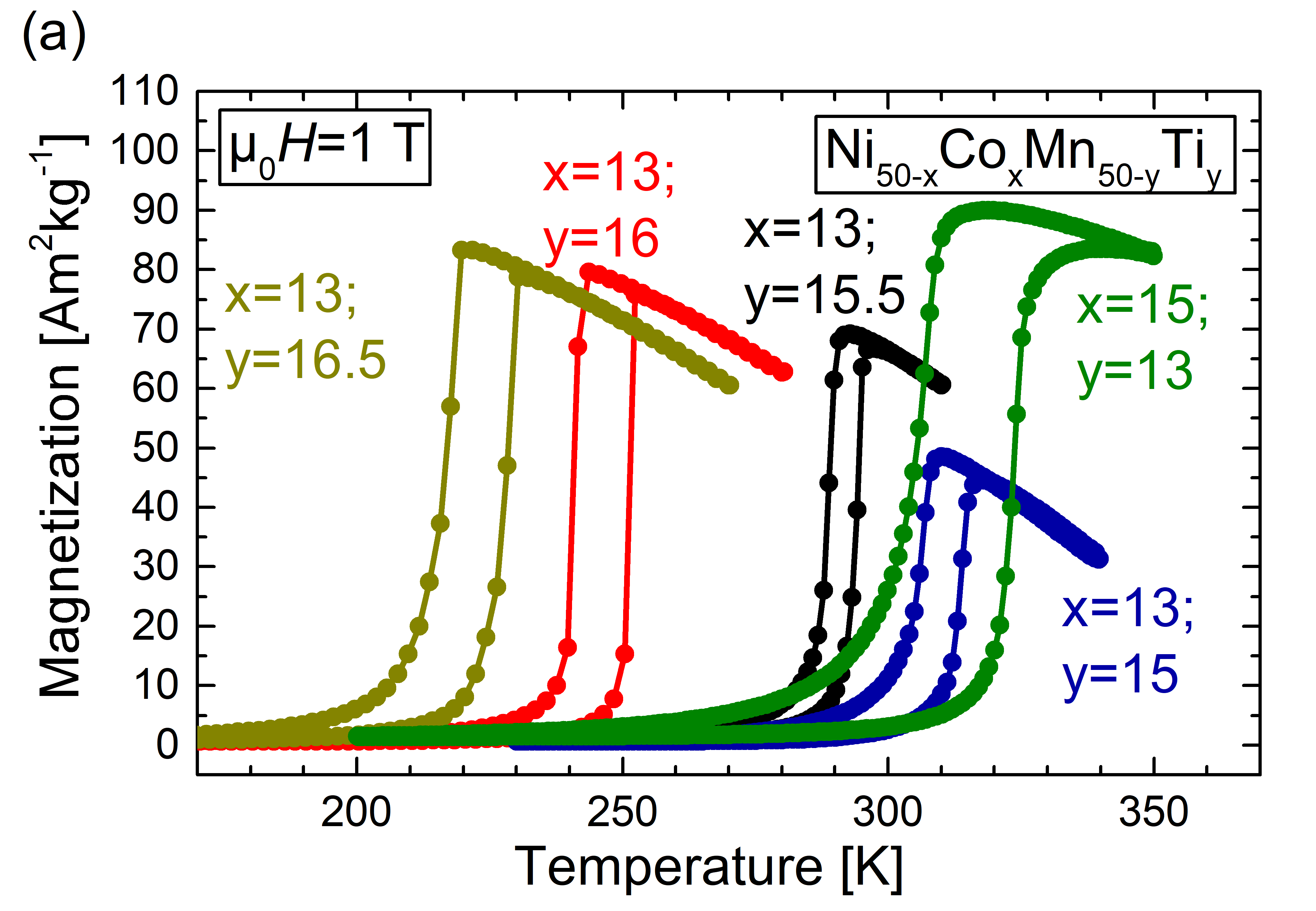}
          \end{minipage}
    \begin{minipage}{0.95\linewidth}
        \centering
        \includegraphics[width=\linewidth]{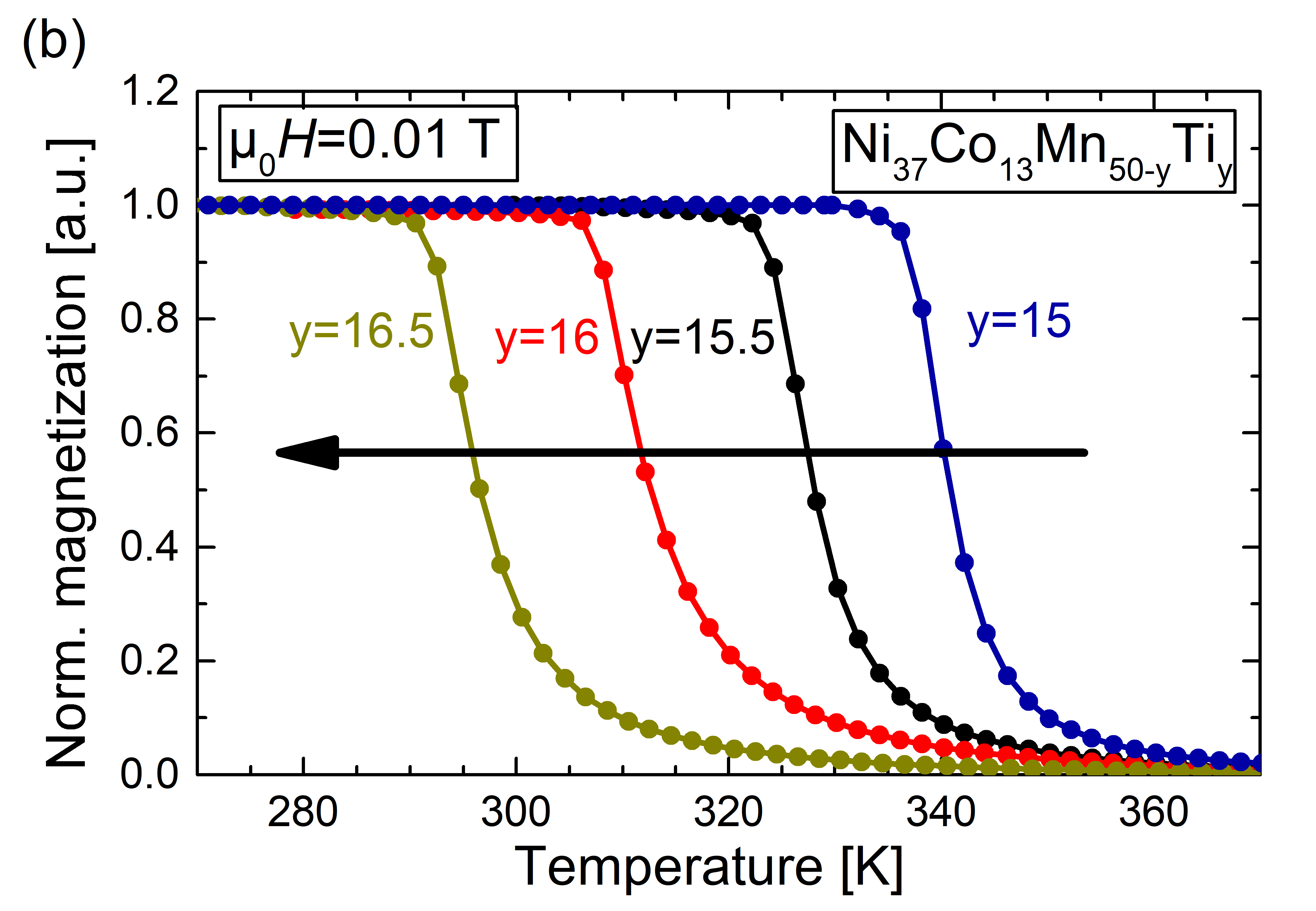}
	\end{minipage}
	\caption{Temperature-dependent magnetization curves for the series of Ni$_\text{50-x}$Co$_\text{x}$Mn$_\text{50-y}$Ti$_\text{y}$ with fixed Co content of \SI{13}{\atom\percent} and one sample with $x=15$ and $y=13$ showing the martensitic phase transition in \SI{1}{\tesla} for heating and cooling (a) as well as $T_C^A$ in \SI{0.01}{\tesla} for heating protocol (b).}
	\label{MT}
\end{figure}

\begin{figure}[!h]
    \centering
    \begin{minipage}{0.95\linewidth}
        \centering
        \includegraphics[width=\linewidth]{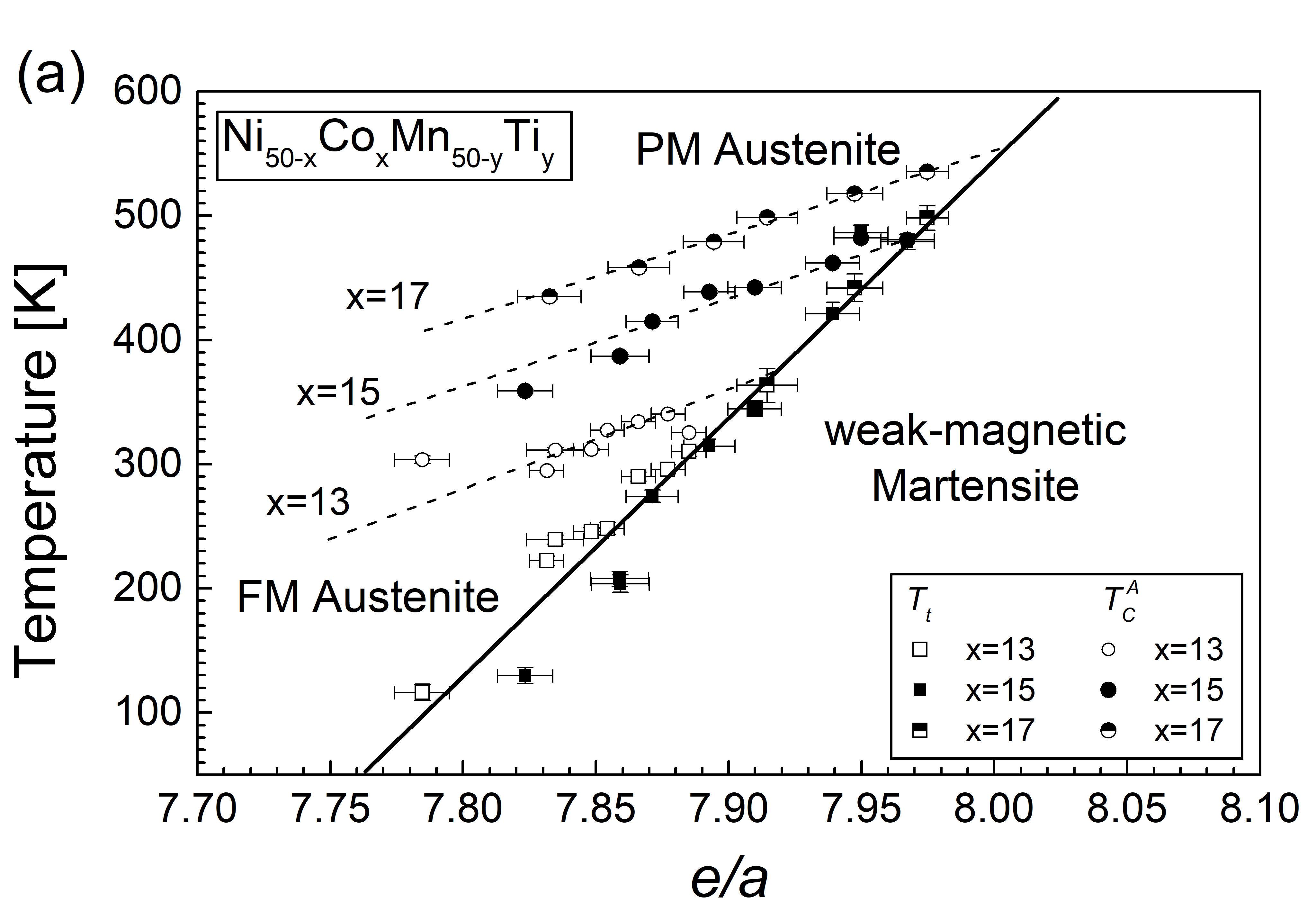}
          \end{minipage}
    \begin{minipage}{0.95\linewidth}
        \centering
        \includegraphics[width=\linewidth]{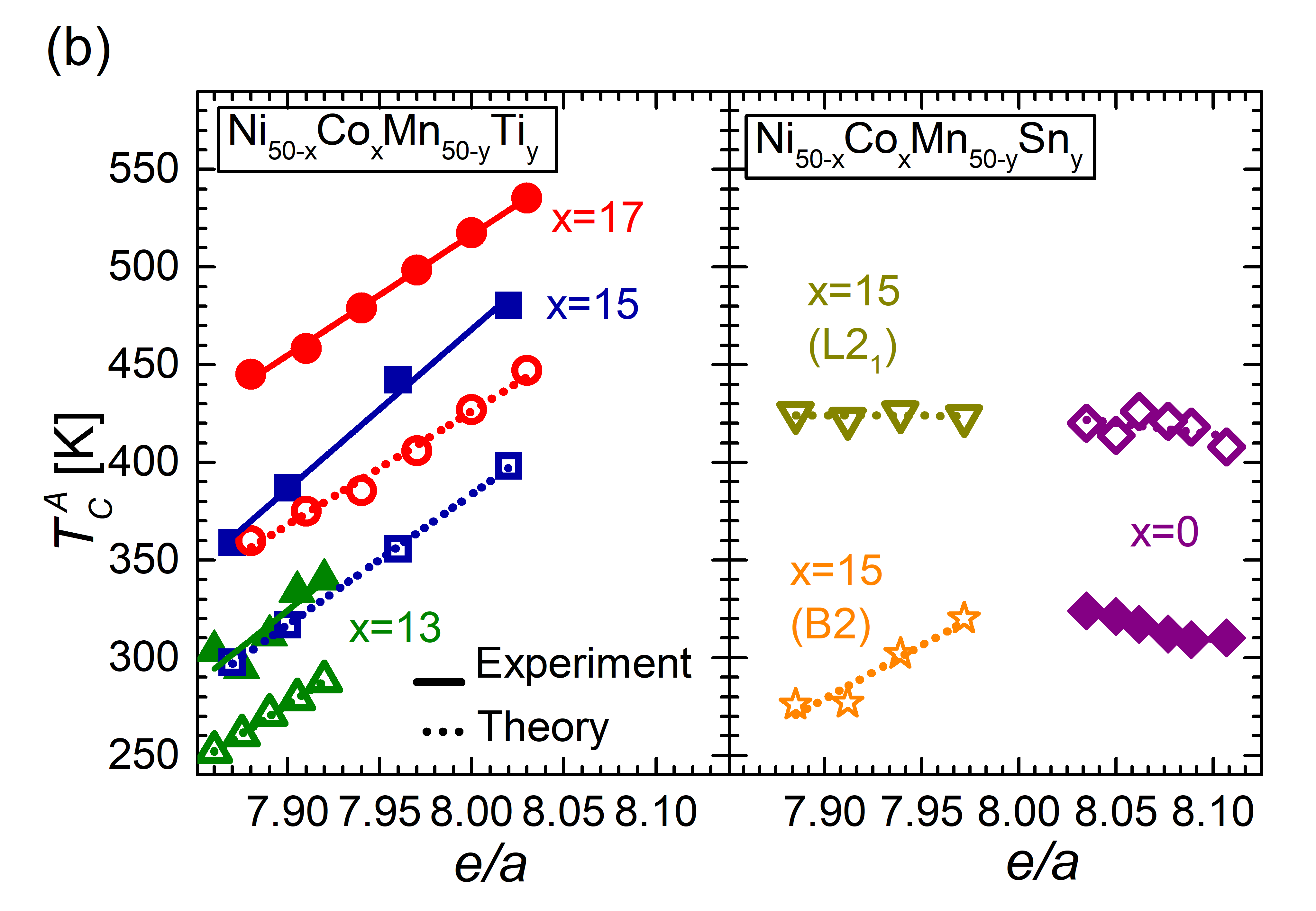}
	\end{minipage}
	\begin{minipage}{0.95\linewidth}
        \centering
        \includegraphics[width=\linewidth]{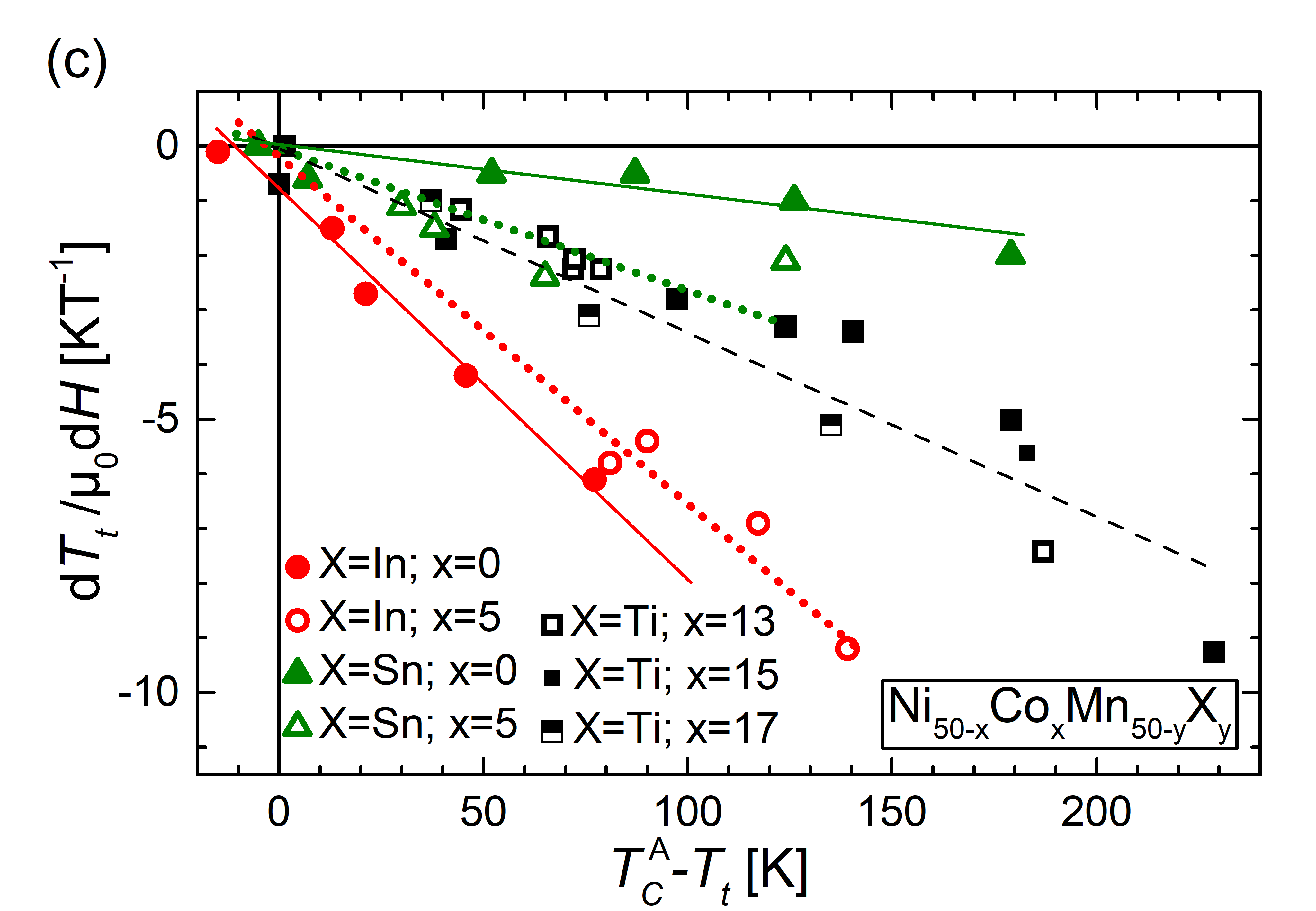}
	\end{minipage}
	\caption{Phase diagram showing the martensitic transition temperatures of all Ni$_\text{50-x}$Co$_\text{x}$Mn$_\text{50-y}$Ti$_\text{y}$ compounds (squares) as well as the austenitic Curie temperatures $T_C^A$ (circles) depending on the $e/a$ ratio (a). The corresponding calculations for a fixed Co content of $x=13$ (green triangles), $x=15$ (blue squares) and $x=17$ (red squares) as well as for the Ni$_\text{50-x}$Co$_\text{x}$Mn$_\text{50-y}$Sn$_\text{y}$ series with $x=0$ (purple diamonds) and $x=15$ (yellow triangles for L2$_1$ ordering and orange stars in the case of B2 disorder) are shown in comparison to the experimental values found for $T_C^A$ in (b). The trend is depicted by linear fits to the data points of the experiment (solid lines) and the calculations (dashed lines). The shift of the transition temperature with respect to an external magnetic field is shown in (c) depending on its difference to the corresponding $T_C^A$ of the same sample for \mbox{Ni-Co-Mn-Ti} (black squares, this work), \mbox{Ni(-Co)-Mn-In} (red circles, ref~\cite{Taubel2018}), and \mbox{Ni(-Co)-Mn-Sn} (green triangles, ref~\cite{Taubel2018}). Fitted lines are drawn here to guide the eye.}
	\label{phase diagram}
\end{figure}

The temperature-dependent magnetization measurements in Fig.~\ref{MT} for samples with varying stoichiometry show that the phase transition temperatures are decreasing with increasing Ti content. This effect is conform with literature~\cite{Wei2015, Wei2016, Neves-Bez2019} and expected from the decreasing $e/a$ ratio, like it is observed for other Ni-Mn-X Heusler systems~\cite{Krenke2006}. The shift of the martensitic transition temperature amounts to an estimate of \SI{50}{\kelvin} per \si{\atom\percent} of Ti. Additionally, $T_C^A$ is decreasing strongly for increasing Ti content (decreasing $e/a$ ratio). Such a significant variation of $T_C^A$ is not present for the \mbox{Ni(-Co)-Mn-X} systems of X=In, Al, Sn, Sb, or Ga, where the Curie temperature is influenced mainly by the Co content but not to this extent by the variation in the Mn-X ratio of the compound~\cite{Sutou2004, Krenke2005, Gottschall2016c}. However, the absolute shift of $T_C^A$ is lower than the shift of the transition temperature. These findings are summarized in the phase diagram including the martensitic transition temperatures as well as the austenitic Curie temperatures for all produced alloys of the Ni$_\text{50-x}$Co$_\text{x}$Mn$_\text{50-y}$Ti$_\text{y}$ system in Fig.~\ref{phase diagram}~(a). The diagram emphasizes that the influence of the stoichiometry (varying Ti and Co content) on the transition temperature can be described uniformly by only considering the $e/a$ ratio of a compound. The behavior of $T_C^A$ is depicted by dashed lines since it is not only depending on $e/a$ but also on the respective Co content of an alloy. Therefore, trend lines exist for constant Co concentrations within the different sample series.

Next, we carried out DFT calculations for a more detailed investigation on the uniquely strong influence of Ti content on $T_C^A$. The comparison between the calculations (open symbols and dashed lines) and the experimentally found data points (closed symbols and solid lines) is depicted in Fig.~\ref{phase diagram}~(b). The theoretical results can reproduce the slope of the $e/a$-dependent $T_C^A$ for the individual  sample series of Co content by using a B2 austenite state for the all-$d$ \mbox{Ni-Co-Mn-Ti} alloys (left panel of Fig.~\ref{phase diagram}~(b)), but with an absolute offset of \SI{90}{\kelvin}, attesting that DFT accurately captures the underlying trends. In order to further understand the behavior of the all-$d$ metal system, we make a comparison with a representative Heusler system with a main group element. Therefore, the experimental data for a \mbox{Ni-Mn-Sn} sample series (purple symbols) is added (right panel of Fig.~\ref{phase diagram}~(b)). The trend fits again to the theoretical model, but with an absolute offset. It is apparent that the evolution of $T_C^A$ with respect to $e/a$ for the all-$d$-metal \mbox{Ni-Co-Mn-Ti} alloy system is much stronger compared to the flat behavior (of yet opposite sign) in the well-studied \mbox{Ni(-Co)-Mn-X} systems with X=Al, In, Sn, Sb~\cite{Kainuma2000, Sutou2004, Krenke2005}. One reason is the increasing amount of Mn on Ti lattice sites (increasing $e/a$) resulting in an enhancement of ferromagnetic interactions $J_{ij}$ between Co and Mn atoms. This stabilizes the ferromagnetic ordering of the austenite phase and leads to an increase in $T_C^A$. In the Ni-Mn-Sn system in contrast, the additional Mn on the D-sites couples antiferromagnetically with the B-site Mn, counteracting with the ferromagnetic interactions between Ni-Mn pairs, which gives rise to a flat decrease of $T_C^A$ with respect to the Mn content. However, Co doping with a concentration as high as \SI{15}{\atom\percent} leads to a diverse effect. A flat behavior of $T_C^A$ over $e/a$ is obtained assuming a L2$_1$ structure for the Sn-based austenite phases, whereas assuming a B2 disorder leads to the same steep variation with decreasing $T_C^A$ for decreasing $e/a$, like it is observed for the Ti-based cases. Therefore, we suspect that the significant change of $T_C^A$ for \mbox{Ni-Co-Mn-Ti} is driven by the B2 disorder. In detail, the different configurations of the Mn-Co interaction for the B2 and L2$_1$ disorder types are responsible for such significant difference in $T_C^A$ with respect to $e/a$. The preferred occurrence of B2 austenite for \mbox{Ni-Co-Mn-Ti} is also in agreement with the reports so far on this material system~\cite{Wei2015, Wei2016, Neves-Bez2019}. 

To further evaluate these correlations, we take a closer look on the evolution of magnetic moments and exchange interactions with changing stoichiometry. Electronic band structure calculations reveal that the moments of Co, Ni and Mn atoms are all lowered with increasing Ti content for \mbox{Ni-Co-Mn-Ti}. By changing the Ti content from \SI{10}{} to \SI{15}{\atom\percent} for a constant Co content of \SI{17}{\atom\percent}, one can see a reduction of \SI{8.7}{\percent}, \SI{16.8}{\percent} and \SI{0.4}{\percent} for Co, Ni and Mn moments, respectively. This results in a decrease of the magnitude of the exchange interactions, particularly for the  ferromagnetic Mn-Co and Mn-Ni interactions (a more detailed discussion can be found in the supplementary material). Both observations of changing moments and exchange interactions consequently enhance the sharp compositional dependence of $T_C^A$ with the Ti content. In addition, the exchange interactions are also influenced by the changing distance between the atoms upon elemental substitutions. The origin of the found behavior for the magnetic moments is tied to the unique feature of the all-$d$-metal Heusler alloys exhibiting a remarkable $d$-$d$ orbital hybridization~\cite{Wei2015}. Such strong hybridization creates sharp peaks with a noticeable pseudo-gap feature of the density of states (DOS) just below the Fermi energy in the minority spin channel (see supplementary, Fig.~S.2). By changing the $e/a$ ratio, also the number of states in the minority spin channel increases faster than the flat and smaller DOS of the majority spin channel, reducing the moments of all magnetic species. A different picture is observed in the \mbox{Ni-Co-Mn-Sn} system, where the Mn and Co/Ni have an opposite trend in moments with Mn being augmented and Co/Ni being reduced (see supplementary for details). Correspondingly, the exchange interactions are reduced only slightly, much less than in the Ti cases. 

The magnetization behavior in Fig.~\ref{MT}~(a) depicts the narrow width of the phase transition for optimally annealed compounds. The difference between martensitic/austenitc start and finish temperatures are in the range of \SIrange{5}{10}{\kelvin} (see table~\ref{tab:Overview}), which is similar to the best reported \mbox{Ni(-Co)-Mn-In} and \mbox{Ni(-Co)-Mn-Sn} Heusler alloys~\cite{LiuNature, Gottschall2015, Taubel2018}. We focus here on the exceptionally sharp phase transition of the Ni$_{37}$Co$_{13}$Mn$_{34}$Ti$_{16}$ sample, that transforms from martensite to austenite state (and vice versa) within \SI{5}{\kelvin}. The thermal hysteresis for samples with \SI{13}{\atom\percent} of Co amounts to \SIrange{8}{12}{K}, which is better than for Ni-Mn-Sn and similar to \mbox{Ni-Co-Mn-Sn} and \mbox{Ni(-Co)-Mn-In} by comparing samples with similar $T_t$ - which is a significant factor for the thermal hysteresis width~\cite{Taubel2018}. Besides a sharp phase transition and a narrow thermal hysteresis, it is important to induce the transition by an external stimulus. Therefore, the shift of the phase transition with an externally applied magnetic field $\frac{dT_t}{\mu_0dH}$ is crucial to achieve a significant magnetocaloric effect for the first field application as well as to overcome the thermal hysteresis for cyclic conditions~\cite{Gutfleisch2016, Scheibel2018, Taubel2018}. The sensitivity of the magnetostructural phase transition to applied magnetic fields is estimated to be \SI{-1.2}{\kelvin\per\tesla} for Ni$_{37}$Co$_{13}$Mn$_{35}$Ti$_{15}$, which shows a transition close to room temperature. For transitions at lower temperatures (by maintaining the Co content of the compound), the field sensitivity is increased to \SI{-2.1}{\kelvin\per\tesla}. 

As already known from other \mbox{Ni(-Co)-Mn-X} systems, $\frac{dT_t}{\mu_0dH}$ is dependent on the martensitic transition temperature and gets larger at lower temperatures~\cite{Taubel2018}. This trend is also observed here: Within all series of constant Co content, the absolute value of $\frac{dT_t}{\mu_0dH}$ increases for increasing Ti content. This behavior is due to the larger magnetization change of the phase transition for reduced transition temperatures. Plotting all data points for $\frac{dT_t}{\mu_0dH}$ depending on the difference of $T_t$ from $T_C^A$ in Fig.~\ref{phase diagram}~(c) shows that this difference is mainly influencing the field sensitivity. The emerging magnetic entropy with decreasing temperature from $T_C^A$ on~\cite{Kihara2014} as driving force for a magnetic-field-induced phase transition is responsible for this effect. This holds true universally regardless of the stoichiometry. By assuming a linear dependence, the estimated slope for the Ti-Heusler system amounts to \SI{0.033}{\per\tesla}. This value is giving a sensitivity for the design of a strong $\frac{dT_t}{\mu_0dH}$ by changing the transition temperature - being equivalent to changing the composition ($e/a$ ratio). For \mbox{Ni(-Co)-Mn-In} alloys, a divergent behavior has been observed~\cite{Gottschall2016c}. The plot in Fig.~\ref{phase diagram}~(c) resembles a linear dependence, which is an approximation for small values of $T_C^A$-$T_t$. A comparison with \mbox{Ni(-Co)-Mn-In} points out that the supposed linear region covers a larger temperature range indicating a lower compensation point~\cite{Gottschall2016c}. Adding the values for other \mbox{Ni(-Co)-Mn-X} systems taken from ref~\cite{Taubel2018} demonstrates that the trend is comparable but the slope indicating the change of $\frac{dT_t}{\mu_0dH}$ with $T_C^A$-$T_t$ varies strongly. As a result, the magnetic field sensitivity of the transition temperature can be tuned more sensitively for certain \mbox{Ni(-Co)-Mn-X} systems. The \mbox{Ni(-Co)-Mn-In} system shows the strongest sensitivity (slope of \SI{0.063}{\per\tesla}), which means that $\frac{dT_t}{\mu_0dH}$ is influenced stronger by a change of the transition temperature itself or by the change of $T_C^A$ than it is the case for \mbox{Ni-Co-Mn-Ti} and \mbox{Ni(-Co)-Mn-Sn} (slope of \SIrange{0.009}{0.026}{\per\tesla}).

Thus, the cyclic response for the magnetic-field induced magnetocaloric effect can be influenced drastically by tuning $T_t$ as well as $T_C^A$ of the alloy. $T_C^A$ must be increased significantly for designing a magnetocaloric Heusler alloy working around room temperature. The purpose here is to shift $T_C$ as far away as possible from the martensitic phase transition temperature. Reaching this goal for \mbox{Ni-Co-Mn-Ti} alloys by raising the Co content leads to a broadening of the transition width and an increased thermal hysteresis (see Fig.~\ref{MT}~(a) and Tab.~\ref{tab:Overview}), which is in agreement with previous studies~\cite{Wei2015, Neves-Bez2019}. This increased thermal hysteresis is commonly regarded as a disadvantage for the reversibility of a magnetocaloric cooling cycle~\cite{Scheibel2018}. However, we report recently that it can be turned into an advantage by using a large thermal hysteresis to prevent the back transformation of the material upon field removal. The initial state can then be reached again by pressure application to the now multicaloric material~\cite{Gottschall2018}.

\subsection{Influence of the heat treatment on the microstructure}
\label{chap:Microstructure}

\begin{figure}[htb]
\begin{center}
\includegraphics[width=\linewidth]{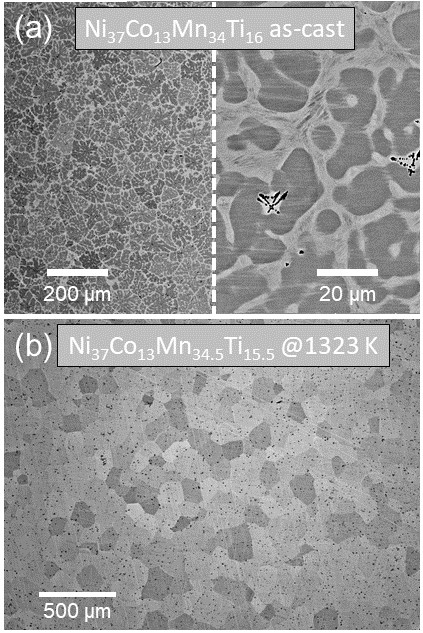}
\caption{SEM images of the as-cast structure of Ni$_{37}$Co$_{13}$Mn$_{34}$Ti$_{16}$ by BSE imaging (a) and of Ni$_{37}$Co$_{13}$Mn$_{34.5}$Ti$_{15.5}$ after ideal annealing at \SI{1323}{\kelvin} showing the resulting grain structure (b).}
\label{Microstructure2}
\end{center}
\end{figure}

In this section, the evolution of the microstructure will be investigated for different heat treatments to examine the differences and correlate the findings to the observed magnetization behavior. The BSE contrast for the as-cast state shown in Fig.~\ref{Microstructure2}~(a) shows two phases - one Ti-rich (dark) and one Mn-rich (bright) compared to the nominal composition (Ni$_{37}$Co$_{13}$Mn$_{33.5}$Ti$_{16.5}$). The averaged composition of the Ti-rich phase is Ni$_{35.2}$Co$_{14.4}$Mn$_{30.6}$Ti$_{19.8}$, whereas the Mn-rich phase consists of Ni$_{36.3}$Co$_{11.0}$Mn$_{39.8}$Ti$_{12.9}$. Therefore, two (off-stoichiometric) Heusler phases with a compositional gradient of different Mn-Ti ratio in between are formed during solidification from the melt. The larger magnification reveals that martensitic structures are visible in the Mn-rich phase even though the overall transition temperature of the nominal composition is around \SI{250}{\kelvin}. This deviation is due to the locally varied transition temperature by the large Mn content. The overall as-cast microstructure is not of columnar but of globular appearance even though the direction of solidification lies in the image plane. This is evidence of slow solidification kinetics.

The typical grain size of the annealed \mbox{Ni-Co-Mn-Ti} samples is in the range of \SIrange{50}{300}{\micro\meter}, as shown representatively in Fig.~\ref{Microstructure2}~(b). In accordance to the observations from the as-cast state, the grains have a globular shape, which deviates from the microstructure observed in \mbox{Ni-Co-Mn-In} Heusler alloys indicating a different solidification behavior~\cite{Liu2009}. The observed grain size is significantly smaller than for conventional \mbox{Ni(-Co)-Mn-X} Heusler alloys, where the grains are in the range of several hundreds of \si{\micro\meter} and up to the millimeter-range. This grain refinement is important for the reported enhanced mechanical strength~\cite{Wei2015} as it is well-known that the yield strength of a metal is inversely proportional to the average grain size diameter after the relation of Hall-Petch~\cite{Hall1951, Petch1953}. The evolution of the strength of this material system in comparison to different microstructures will be reported in more detail in a separate study.

\begin{figure*}[]
\begin{center}
\includegraphics[height=1.25\textwidth]{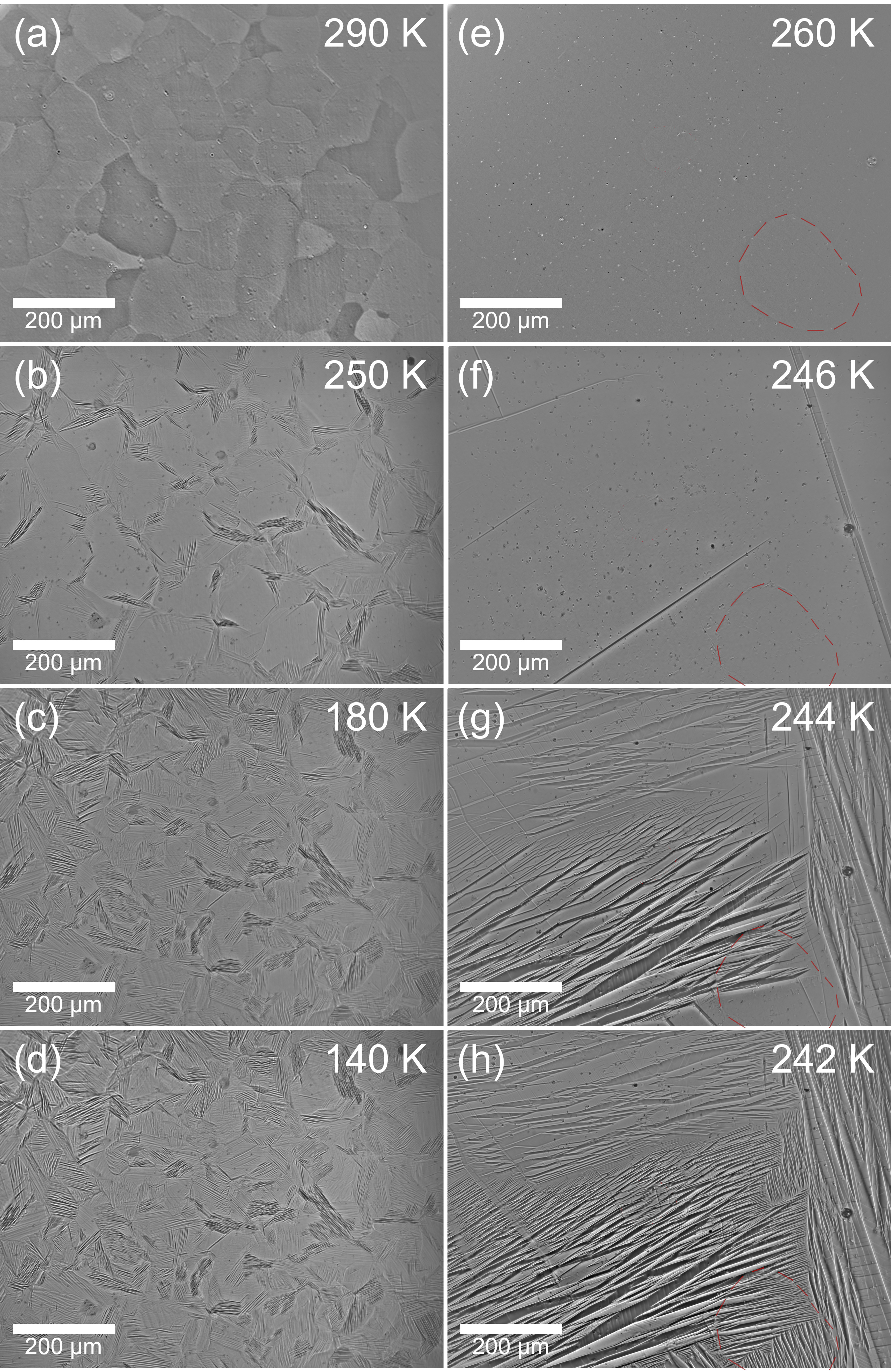}
\caption{Temperature-dependent microstructure images by optical microscopy of Ni$_{37}$Co$_{13}$Mn$_{34}$Ti$_{16}$ for different heat treatments. The non-ideal sample annealed at \SI{1173}{\kelvin} is shown in (a)-(d) and the optimized sample annealed at \SI{1323}{\kelvin} in (e)-(h). The dashed red line indicates a small grain embedded in the abnormally large grown grain orientation in (e)-(h).}
\label{Microstructure}
\end{center}
\end{figure*}

Temperature-dependent microstructural images comparing the martensitic phase transition for a non-ideal (\SI{1173}{\kelvin}) and an optimized (\SI{1323}{\kelvin}) annealing temperature of the same parent as-cast sample (Ni$_{37}$Co$_{13}$Mn$_{34}$Ti$_{16}$, see Fig.~\ref{Microstructure2}~(a) for the corresponding SEM image) are shown in Fig.~\ref{Microstructure}. In-situ heating and cooling videos of the polarized light microscopy studies can be found in the supplementary material. The crystal structure of the formed martensite has been determined by X-ray diffraction experiments of polycrystalline powder. Since the preparation requires crushing of the bulk sample into powder followed by a heat treatment, the analyzed diffraction pattern is not assumed to resemble the bulk state necessarily. The resulting pattern suggests a 5M modulated martensite structure with the lattice parameters $a=$\SI{4.35}{\angstrom}, $b=$\SI{5.45}{\angstrom} and $c=$\SI{4.25}{\angstrom} and an angle of $\beta=$\SI{93.5}{\degree}. This result is in good agreement with literature analyses for Ni-Co-Mn-Ti samples reporting 5M structures \cite{Wei2015, Liu2019, Liu2019a}. We emphasize that the preparation procedure may have influenced the preferred martensite state and the martensite variants observed by microscopical surface investigations might be of different crystal structure.

Both heat treatments lead to austenitic samples at room temperature ((a) and (e)). Upon cooling of the sample homogenized at \SI{1173}{\kelvin} (Fig.~\ref{Microstructure}~(a)-(d)), the first martensitic structures appear at \SI{285}{\kelvin}. This is in agreement with the magnetization curve showing a first drop of the sample magnetization at the martensitic start temperature of \SI{280}{\kelvin} (see Fig.~\ref{annealing}~(a)). These structures appear at grain boundaries of the austenitic grains and the variants grow with the long axis parallel to the grain boundaries. Upon further cooling, new martensite structures nucleate beside the already present martensite structures towards the center of the grain. At a temperature of \SI{250}{\kelvin}, most of the material in the vicinity of grain boundaries is already transformed to martensite, whereas the center of the grains is still in completely austenitic state (see Fig.~\ref{Microstructure}~(b)). At \SI{180}{\kelvin}, a larger amount of completely martensitic grains is visible and the overall phase ratio of martensite is growing as seen in Fig.~\ref{Microstructure}~(c). At the lowest temperature of \SI{140}{\kelvin}, most of the sample is in martensitic state, only one grain still has residual austenite  present in the center. The final structure of the martensite variants resembles the way how the phase is growing upon cooling: small martensite needles represent the former grain boundaries of the austenite phase. From these needles along the grain boundaries, martensite formation happens step-wise in parallel to the present structures towards the inside of the grain. An analysis by EDS indicates that the reason for this broad phase transition is a variation of the stoichiometry within the sample. After an insufficient heat treatment, the microstructure is still separated in Ti-rich regions and Mn-rich regions. There is no sharp phase boundary but a compositional gradient between the centers of each phase. This compositional difference is less pronounced compared to the as-cast state in Fig.~\ref{Microstructure2}~(a). An annealing at \SI{1173}{\kelvin} is not sufficient to homogenize the sample properly. Consequently, the phase transition from austenite to martensite and vice versa happens over a broad temperature region because the transition temperature is strongly depending on the exact stoichiometry and therefore varies locally.  

In a second step, the martensite formation upon cooling is investigated for the ideal heat treatment at \SI{1323}{\kelvin} for another piece of the same parent as-cast sample (Fig.~\ref{Microstructure}~(e)-(h)). At room temperature, the sample is also completely in austenite state, however, the grain structure is different. Only one small grain is present (encircled in red), embedded in a large grain orientation as seen in Fig.~\ref{Microstructure}~(e). The austenite is stable down to a temperature of \SI{246}{\kelvin}, where the first martensite variants appear (Fig.~\ref{Microstructure}~(f)). These martensite variants are slightly growing upon further cooling and further martensite nucleation happens at \SI{245}{\kelvin}. Subsequently, the phase fraction of martensite is growing rapidly and within a temperature range of \SI{4}{\kelvin} (Fig.~\ref{Microstructure}~(g)+(h)), the whole area is covered with martensite phase at \SI{242}{\kelvin}. This investigation is in perfect agreement with the magnetization curve, where the sharp starting temperature of the martensite formation is determined at \SI{243}{\kelvin} (in a magnetic field of \SI{1}{\tesla}). At \SI{238}{\kelvin} - represented by Fig.~\ref{Microstructure}~(h) - the martensite phase fraction determined from the magnetization of the sample is about \SI{90}{\percent}. Large parallel growing variants are present in the left part of the image. In contrast, the growth of the variants is slightly hindered - but not stopped - at the grain boundary of the small grain visible in the bottom right part of the image, indicating a low-angle grain boundary. The different grain orientation affects the growing direction of the martensite. A direct comparison of the martensite state for the two different heat treatments in Fig.~\ref{Microstructure}~(d) and (h) reveals obvious differences. The martensite nucleation for the non-ideal heat treatment starts at grain boundaries and causes very narrow structures upon further nucleation and growing. Contrary, the ideal annealing leads to a very sharp phase transition resulting in larger martensitic structures because the growth is rarely stopped or hindered due to the huge grain size. 

This investigation shows that a proper homogenization of the sample stoichiometry by an optimized heat treatment is essential for a sharp magnetostructural phase transition in \mbox{Ni-Co-Mn-Ti} Heusler alloys. As a result of an increased annealing temperature, the grain structure can tend to form very large grains because of the provided thermal energy. Such an abnormal grain growth can then lead to grains that are much larger than the typically found grain sizes for the samples of the \mbox{Ni-Co-Mn-Ti} Heusler system in the present study (see Fig.~\ref{Microstructure2}~(b)). These large grains are then in the range of those for other \mbox{Ni-Mn-X} Heusler alloys. The untypical grain structure leads to an exceptionally sharp phase transition (see red curve in Fig.~\ref{MT}~(a)), which can be a result of the reduced amount of grain boundaries that can act as a barrier for the growth of the martensitic phase.

\subsection{Measurement of the magnetocaloric effect by $\Delta s_T$ and $\Delta T_{ad}$}
\label{chap:Magnetocaloric}

\begin{figure}[htb]
    \centering
    \begin{minipage}{0.95\linewidth}
        \centering
        \includegraphics[width=\linewidth]{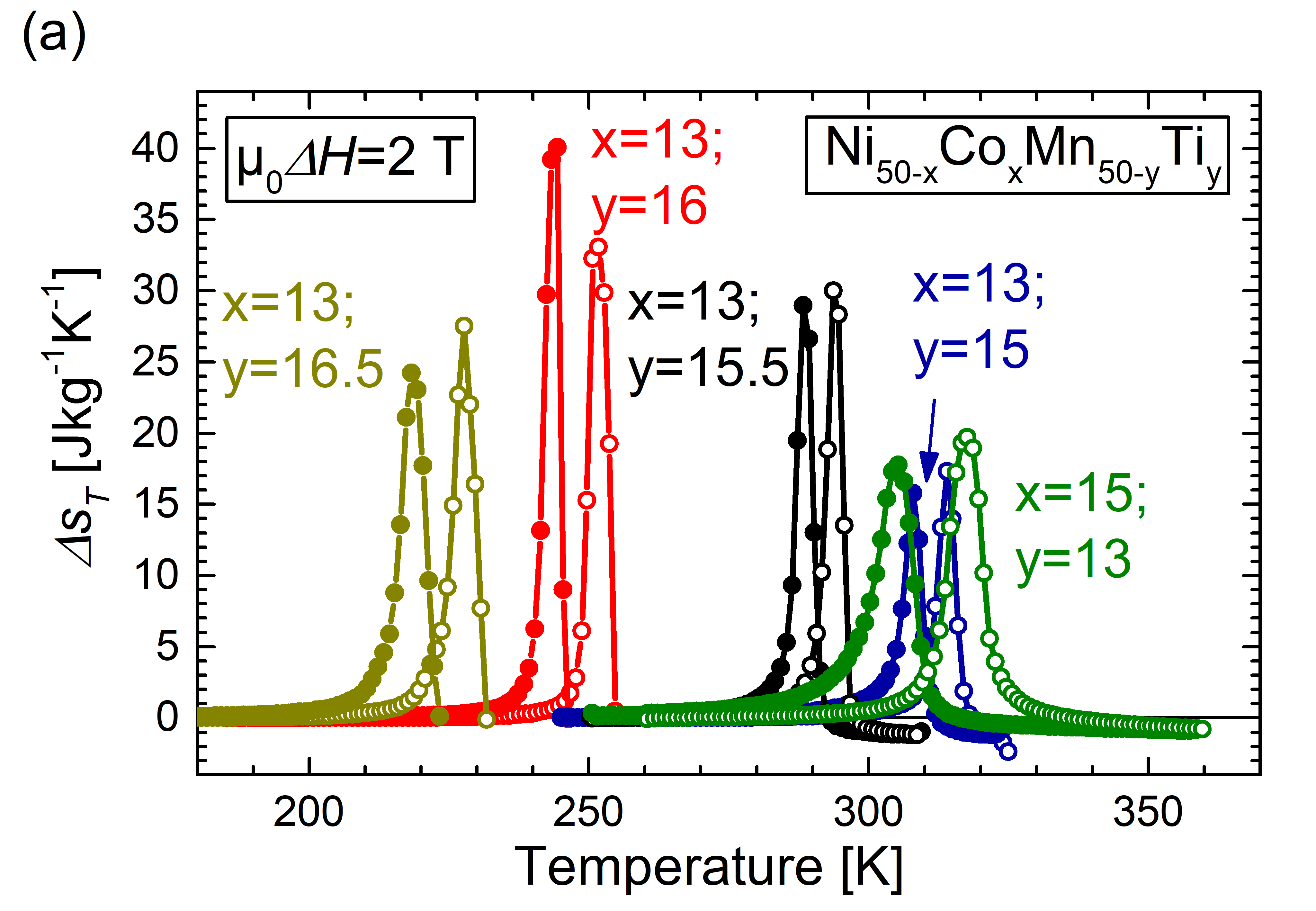}
          \end{minipage}
    \begin{minipage}{0.95\linewidth}
        \centering
        \includegraphics[width=\linewidth]{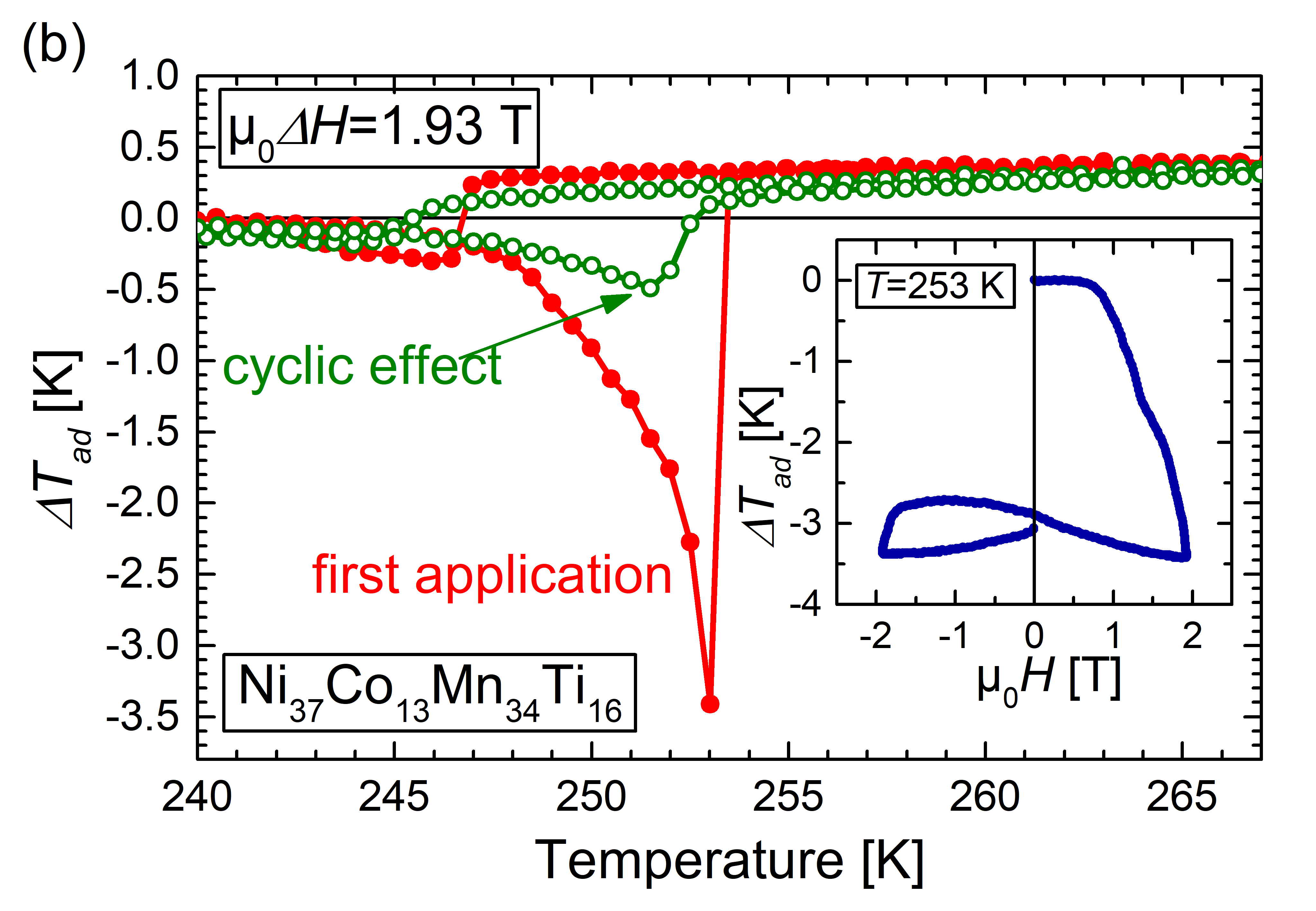}
	\end{minipage}
	\caption{$\Delta s_{T}$ curves for the series of Ni$_\text{50-x}$Co$_\text{x}$Mn$_\text{50-y}$Ti$_\text{y}$ with fixed Co content of \SI{13}{\atom\percent} and one sample with $x=15$ and $y=13$, determined from temperature-dependent magnetization measurements (isofield protocol) upon heating (open symbols) and cooling (full symbols) for magnetic field changes of \SI{2}{\tesla} (a). Directly measured adiabatic temperature change for the Ni$_{37}$Co$_{13}$Mn$_{34}$Ti$_{16}$ sample in continuous mode for the first field application (red full symbols) and for the reversible effect upon the second magnetic-field cycle (green open symbols) for field changes of \SI{2}{\tesla} (b). The magnetic field dependent $\Delta T_{ad}$ at the starting temperature of \SI{253}{\kelvin} for the first two magnetic-field cycles is shown in the inset.}
	\label{DeltaS}
\end{figure}

To assess the potential of the magnetocaloric \mbox{Ni-Co-Mn-Ti} compounds for cooling applications, the isothermal entropy change $\Delta s_T$ as well as the adiabatic temperature change $\Delta T_{ad}$ are evaluated. The isothermal entropy change is determined for magnetic field changes of \SI{2}{\tesla} and shown in Fig.~\ref{DeltaS}~(a) (compare respective $M(T)$-curves in Fig.~\ref{MT}~(a)). Due to the exceptionally sharp phase transition, the Ni$_{37}$Co$_{13}$Mn$_{34}$Ti$_{16}$ compound provides the largest $\Delta s_T$ for this series. It has been determined from isofield $M(T)$ measurements to avoid overestimation~\cite{Caron2009} and amounts to \SI{38}{\joule\per\kilo\gram\per\kelvin} in a magnetic field change of \SI{2}{\tesla} at a transition temperature of around \SI{250}{\kelvin}. This is a giant value for a bulk Heusler alloy at this field change, which is significantly larger compared to the magnetocaloric $\Delta s_T$ determined from magnetization measurements for the all-$d$-metal Heusler alloys in literature~\cite{Wei2015, Wei2016, Neves-Bez2019}. The Ni$_{37}$Co$_{13}$Mn$_{34.5}$Ti$_{15.5}$ compound with a transition around room temperature shows a $\Delta s_T$ of around \SI{30}{\joule\per\kilo\gram\per\kelvin}. Despite the broader phase transition, the Ni$_{35}$Co$_{15}$Mn$_{37}$Ti$_{13}$ sample still shows a large $\Delta s_T$ of around \SI{20}{\joule\per\kilo\gram\per\kelvin}, which is comparable to the one of the Ni$_{37}$Co$_{13}$Mn$_{35}$Ti$_{15}$ sample with a similar transition temperature.

The sharp peak shape indicates that the maximum possible $\Delta s_T$ is not reached in \SI{2}{\tesla} and larger fields would be necessary to obtain the plateau of the saturated $\Delta s_T$. Other works show that this saturation is not even observed for field changes of \SI{5}{\tesla}~\cite{Wei2015}. Considering the magnetization change $\Delta M$ of \SI{75}{\ampere\meter\squared\per\kilo\gram} and a shift of the transition $\frac{dT_t}{\mu_0dH}$ of \SI{-1.6}{\kelvin\per\tesla} (for Co content of \SI{13}{\atom\percent} at room temperature), an estimation of the maximum possible isothermal entropy change by the Clausius-Clapeyron approximation would give $\Delta s_T^{max}=\frac{\Delta M}{\frac{dT_t}{\mu_0dH}}=\SI{47}{\joule\per\kilo\gram\per\kelvin}$. 

In Ni-Mn-In alloys, depending on the transition temperature, a complete transition can be reached in \SI{2}{\tesla} for some compounds because of the large $\frac{dT_t}{\mu_0dH}$~\cite{Gottschall2016c, Taubel2018}. For the \mbox{Ni-Co-Mn-Ti} alloys, $\frac{dT_t}{\mu_0dH}$ is significantly smaller leading in combination with a very sharp transition to the observed large entropy changes.

The adiabatic temperature change for the Ni$_{37}$Co$_{13}$Mn$_{34}$Ti$_{16}$ sample with the largest $\Delta s_T$  is shown in Fig.~\ref{DeltaS}~(b). The maximum $\Delta T_{ad}$ of \SI{-3.5}{\kelvin} for the first field application can be obtained at a temperature of \SI{253}{\kelvin}. The evolution of $\Delta T_{ad}$ over the applied magnetic field strength shown in the inset indicates also here that the transition is not completed in \SI{2}{\tesla}. At the largest field value, the temperature is still decreasing with considerable slope, which stops immediately when the field is being removed again. Due to the reduced $\frac{dT_t}{\mu_0dH}$ compared to other \mbox{Ni(-Co)-Mn-X} systems~\cite{Gottschall2015}, a field change of \SI{2}{\tesla} cannot shift the transition temperature enough to fully induce the austenite state. The decreasing sample temperature due to the negative adiabatic temperature change of the inverse magnetocaloric effect works against the austenite formation of the field-induced phase transition. Therefore, the temperature shift per field unit multiplied by the applied magnetic field ($\frac{dT_t}{\mu_0dH}\cdot \mu_0\Delta H$) gives the maximum possible MCE~\cite{Skokov2012}. To reach the maximum $\Delta T_{ad}$, this must be equal or larger than the sum of the transition width and the adiabatic temperature change for the corresponding field change at a certain starting temperature. Even though the transition width is larger, the Ni$_{35}$Co$_{15}$Mn$_{37}$Ti$_{13}$ compound with higher Co content and thus a larger $\frac{dT_t}{\mu_0dH}$ of \SI{-2.8}{\kelvin\per\tesla} yields a maximum $\Delta T_{ad}$ of \SI{-4}{\kelvin} for the first field application.

This compound shows a low cyclic response of the material upon the second field application step, which illustrates the low reversibility of the magnetocaloric effect. Since a magnetic field of \SI{2}{\tesla} shifts the transition temperature around \SI{3}{\kelvin} to lower temperatures, the thermal hysteresis of around \SI{10}{\kelvin} cannot be overcome. Nevertheless, a certain temperature change of around \SI{0.5}{\kelvin} is present for the second field cycle (green open symbols in Fig.~\ref{DeltaS}~(b) and negative fields in the inset), which is due to the lowered energetic barrier that results from the not completed first transition cycle leading to minor loops of thermal hysteresis~\cite{Gottschall2015}.

\section{Conclusions}
\label{Conclusions}

We present a systematic study on the optimization of the heat treatment procedure for improved magnetocaloric properties of Ni$_\text{50-x}$Co$_\text{x}$Mn$_\text{50-y}$Ti$_\text{y}$ all-$d$-metal Heusler alloys. With an annealing step at \SI{1323}{\kelvin} for \SI{96}{\hour} followed by rapid water quenching, very sharp martensitic phase transitions can be achieved with isothermal entropy changes of up to \SI{38}{\joule\per\kilo\gram\per\kelvin} in a magnetic field change of \SI{2}{\tesla} for Ni$_{37}$Co$_{13}$Mn$_{34}$Ti$_{16}$ with a transition temperature of around \SI{250}{\kelvin}. Microstructural investigations show that lower heat treatment temperatures do not lead to a homogeneous distribution of the stoichiometric variances present in the as-cast structure. An optimized annealing procedure results in a homogeneous distribution and the sharpest phase transition can be obtained for irregularly large grains. 

Using the optimal annealing procedure, a phase diagram including $T_t$ and $T_C^A$ is created by varying the Co and Ti content systematically. The behavior of $T_C^A$ is found to be different from Ni-(-Co)-Mn-X Heusler alloys with X being a main group element. Theoretical calculations are in very good agreement with the experimental data and point out that the origin of this trend is due to the preferred B2 disorder for the \mbox{Ni-Co-Mn-Ti} alloys. This leads to an enhancement of the ferromagnetic coupling between Co-Mn moments, which stabilizes the FM phases upon increasing Mn concentration in the alloys. The direct comparison of the DOS for \mbox{Ni-Co-Mn-Ti} with that of \mbox{Ni-Co-Mn-Sn} reveals a pseudo gap feature for the Ti-based alloys with the Fermi level located on the slope, which is a result of the strong hybridization in the all-$d$-metal Heusler alloys.

The value of the adiabatic temperature change measured in this study for Ni$_{37}$Co$_{13}$Mn$_{34}$Ti$_{16}$ amounts to \SI{-3.5}{\kelvin} for the first field application and \SI{0.5}{\kelvin} reversibly for further field cycles. Even though this compound shows a very sharp phase transition, magnetic fields of \SI{2}{\tesla} are too low to completely induce the phase transition. The reason is the low shift of the transition temperature in applied magnetic fields ($\frac{dT_t}{\mu_0dH}$ for the series with a CO content of \SI{13}{\atom\percent} is around \SIrange{-1}{-2}{\kelvin\per\tesla}). In order to design an improved magnetic-field induced magnetocaloric performance, variation in the Co-content shows a general trend for $\frac{dT_t}{\mu_0dH}$ being directly depending on the difference between the transition temperature and the Curie temperature of the austenite. Thus, a larger field sensitivity at a constant transition temperature can be achieved by increasing the Co content of the alloy. However, this results in a broadening of both the phase transition and the thermal hysteresis. For Ni$_{35}$Co$_{15}$Mn$_{37}$Ti$_{13}$, a phase transition can be established at room temperature with large magnetization changes, an isothermal entropy change of \SI{20}{\joule\per\kilo\gram\per\kelvin} and an adiabatic temperature change of \SI{-4}{\kelvin} for a magnetic field change of \SI{2}{\tesla}. The increased hysteresis for larger Co contents is usually a drawback for cyclic applications, which we aim to turn into an advantage by using it together with a mechanical field in a multi-stimuli cycle. 

\section*{Acknowledgments}
We acknowledge the financial support by the European Research Council (ERC) under the European Unions Horizon 2020 research and innovation programme (Grant No. 743116 - project "Cool Innov"), by the Deutsche Forschungsgemeinschaft (DFG, German Research Foundation) over the CRC "HoMMage" Project-ID 405553726 – TRR 270, by the Helmholtz Association via the Helmholtz-RSF Joint Research Group with the Project No. HRSF-0045, and by the HLD at HZDR, a member of the European Magnetic Field Laboratory (EMFL). 

The authors thank David Koch (TU Darmstadt) for the measurement and refinement of XRD patterns and the valuable discussions about the martensite structure analysis.


\bibliography{Manuscript_Taubel_accepted}

\end{document}